\documentclass[12pt]{article}

\usepackage{array} 
\usepackage{epsfig}
\usepackage{amssymb}
\usepackage{amsmath}                       
\usepackage{graphics,graphpap}
\usepackage{graphicx}
\usepackage{epstopdf}

\renewcommand{\thefootnote}{\fnsymbol{footnote}}

\begin{document}

\begin{titlepage}

\vskip1.5cm
\begin{center}

{\Large \bf \boldmath Formalism for Simulation-based Optimization of Measurement Errors 
in High Energy Physics }

    \vskip1.3cm {
     Yuehong Xie\footnote{Yuehong.Xie@cern.ch}
\vskip0.3cm    
 {\em  University of Edinburgh, Edinburgh EH9 3JZ, United Kingdom }}
    \vskip0.5cm 


\vskip5cm

{\large\bf Abstract:\\[10pt]} \parbox[t]{\textwidth}{

Miminizing errors of the physical parameters of interest should be the ultimate
goal of any event selection optimization in high energy physics data analysis 
involving parameter determination. Quick and reliable error estimation is a crucial ingredient  for realizing this goal. 
In this paper we derive a formalism for direct evaluation of measurement  errors 
using the signal probability density function and large fully simulated signal and background samples
without need for  data fitting and background modelling. 
We illustrate the elegance of the formalism 
 in the case of event selection optimization for CP violation measurement in B decays. 
The implication of this formalism on choosing event variables for data analysis  is discussed.

}

\vfill

\end{center}
\end{titlepage}

\setcounter{footnote}{0}
\renewcommand{\thefootnote}{\arabic{footnote}}

\newpage

\section{Introduction}

While performing event selection optimization has become a common practice in high energy physics data analysis, a not so 
well known fact is that the optimization problem is often poorly formulated. Many efforts have been made in developing
complicated techniques~\cite{statpat,tmva}
to separate signal from background in the multi-dimensional space of event variables, 
but much less attention is paid to the more 
fundamental questions - what is the goal of event selection optimization and how can we quantify it?

In principle, event selection optimization in analysis measuring particle properties  should be aimed for the true purpose of 
the analysis, i.e. to produce the  most accurate and precise physics results~\cite{open}. In practice we have to refine this goal 
to make it  quantifiable. First of all, it is hardly possible to quantify change of  systematic uncertainties with regard to
change of  selection criteria in the optimization stage. Concerning event selection, the requirement to control systematic 
uncertainties is supposed to be qualitatively fulfilled by imposing constraints on the allowed selection criteria
and  choices of fitting variables.
The selection optimization problem effectively becomes the problem of finding the 
selection criteria that minimize
statistical uncertainties on the  parameters of interest under necessary constraints.
Unless otherwise stated, we will use the word error to refer to statistical uncertainty in this paper.
Secondly, optimization is usually implemented by maximizing or minimizing a single objective function.
In case several physical parameters are  measured in one analysis, there are different  strategies to
define the objective function, for example  by focusing on  the most important parameter, by using the weighted linear 
sum of different objective functions, or by using the determinant of the  covariance matrix of all the 
parameters of interest. A fully determined covariance matrix is sufficient for evaluation of all measurement errors
and  any  form  of objective functions. 

In a usual analysis, a mathematical model describing both signal and background events 
needs to be identified first. Then statistical techniques (see e.g.~\cite{Lyons} and~\cite{Cowan})
such as  the maximum likelihood method are used to fit the model to some data 
and find the parameter estimates. Finally the parameter covariance matrix  is evaluated  using the parameter estimates,
for example by evaluating the Hessian matrix of the log-likelihood function at the the maximum likelihood solution point.
While in principle guaranteed to work and ``straightforward" to implement, this procedure is very inefficient or even unfeasible  for
optimization tasks which require to iterate the procedure many times to find the optimal selection criteria.
It is difficult to dynamically identify a  model that accurately describes the criteria-dependent background.
Even if such a model can be found, the time needed to do the large number of data fitting
is usually beyond affordability. A clever way to perform optimization should be sought. The crucial 
ingredient of event selection optimization is a quick and reliable estimation of the parameter errors ( or the 
covariance matrix).

\section{ Error estimation  in simulation-based optimization }
\label{sec:method}

Most often event selection optimization is pursued using large samples of fully simulated 
signal and background events. Therefore the true values of the physical parameters used
to generate the simulation data are already known.
We can skip the fitting step and directly evaluate what the errors of these parameters 
would be if  they were estimated by fitting a model to the data. The errors obtained this way
are minimized in the optimization process.

\subsection{ The general formalism}

Suppose we have a selected data sample of both signal and background events 
each consisting of a set of $k$ measured quantities $X = (x_1,...,x_k)$. 
Let $P_s (X; \Theta)$ and $P_b (X)$ be the probability density function (p.d.f.) describing 
the signal and background events respectively, where $\Theta = (\theta_1,...,\theta_m)$ 
is a set of parameters whose true values are unknown in real data analysis
but  accessible in simulation-based optimization.
  $\Theta$ includes the  physical parameters we are interested in
and  maybe also some signal-related experimental parameters that need to be determined from the fit. 
Here we have assumed that the background p.d.f. is totally fixed using sidebands, control channels
or full simulation and thus has no fit parameters.

The p.d.f. describing the total sample can be written as 
\begin{equation}
\label{equ:Pt}
 P_t (X;\Theta,F) = F\cdot P_s (X;\Theta) + (1-F)\cdot P_b (X)
\end{equation}
where $F$ is the fraction of signal events in the sample, called  the global purity.
This kind of parameter estimation problem is usually solved by maximizing the likelihood function
\begin{equation}
\label{equ:L}
 L(\Theta, F) \equiv   \prod_{l=1}^{N_s+N_b} P_t(X_l;\Theta,F)
\end{equation}
where $N_s (N_b)$ is the number of signal (background) events in the selected sample. 

In most high energy physics data analyses $F$ and $\Theta$ are determined using
different observables, resulting in  weak and negligible correlations between them.
Using this fact, we can approximate
the inverse  $V^{-1}$ of the covariance matrix $V_{ij}= cov[\hat{\theta_i},\hat{\theta_j}]$ using
the Hessian matrix of the log-likelihood function with respect to $\Theta$ only:
\begin{equation}
  ( V^{-1})_{ij} = H_{ij}= - \frac{\partial^2 \ln L}{\partial \theta_i \partial \theta_j}
   =  - \sum_{l=1}^{N_s + N_b} \frac{\partial^2 \ln P_t (X_l;\Theta,F)} {\partial \theta_i \theta_j}.
\end{equation}
The Hessian matrix should be evaluated at the true values of the parameters $\Theta$  in principle,
but can only be evaluated at the maximum likelihood solution point in real data analysis. 
Since the true values of all parameters can be known in simulation-based optimization, we have no need to perform
data fitting for evaluation of the Hessian matrix. 
 
Using Eq.~(\ref{equ:Pt}), the Hessian matrix becomes
\begin{equation}
\label{equ:H_full}
( V^{-1})_{ij} =  H_{ij} =  \sum_{l=1}^{N_s + N_b} \left [
(f^2_l-f_l) \cdot  \frac{\partial \ln P_s (X_l;\Theta)} {\partial \theta_i}  \cdot \frac{\partial \ln P_s (X_l;\Theta)} {\partial \theta_j} 
- f_l \cdot \frac{\partial^2 \ln P_s (X_l;\Theta)} {\partial \theta_i \partial \theta_j }
\right ] 
\end{equation}
where 
\begin{equation}  
f_l \equiv F\cdot P_s (X_l;\Theta)/(F\cdot P_s (X_l;\Theta) + (1-F)\cdot P_b (X_l))
\end{equation}
is the local purity in the vicinity of the $l$-th event, which can be estimated  numerically,
for example by counting the number of signal and background events in a zone containing this event. 
 
In the large sample limit, we have
\begin{equation}
 \sum_{l=1}^{N_s + N_b} f_l \cdot G(X_l) = \sum_{l=1}^{N_s} G(X_l)
\end{equation}
for any function $G(X)$. Assuming   this relation holds for moderate sample size,
we can rewrite Eq.~(\ref{equ:H_full}) as
\begin{equation}
\label{equ:master}
( V^{-1})_{ij} =  H_{ij} =  \sum_{l=1}^{N_s } \left [
(f_l-1) \cdot  \frac{\partial \ln P_s (X_l;\Theta)} {\partial \theta_i} \cdot \frac{\partial \ln P_s (X_l;\Theta)} {\partial \theta_j}
-  \frac{\partial^2 \ln P_s (X_l;\Theta)} {\partial \theta_i \partial \theta_j }
\right ] ,
\end{equation}  
which is the master formula for evaluation of the Hessian matrix. An important feature of this result
is that any element of the Hessian matrix   can be expressed as sum of contributions from each signal event
that can be evaluated using the signal p.d.f. and the local purity factor. There is no need to 
model the background as all information  about background is contained in the local purity factor dynamically evaluated from 
fully simulated data. Once the Hessian matrix is estimated, it is trivial to invert it to get the covariance matrix $V$ for the parameters $\Theta$.
Errors of physical parameters and eventually the optimization objective function can be easily evaluated.

\subsection{ A simplified situation }

Very often we care about the error of only one physical parameter, denoted as  $\theta_1$, 
in an analysis and this parameter has weak and
insignificant correlations with other fit parameters. In this case an  objective function to be maximized
can be simply defined as inverse of the variance of the parameter $\theta_1$:
\begin{equation}
 Q = \frac{1}{var(\theta_1)} = H_{11} =   \sum_{l=1}^{N_s } \left [
(f_l-1) \cdot  \left( \frac{\partial \ln P_s (X_l;\Theta)} {\partial \theta_1} \right) ^2
-  \frac{\partial^2 \ln P_s (X_l;\Theta)} {\partial \theta_1 ^2 } 
\right ].
\end{equation}
If the signal p.d.f. is a linear function of $\theta_1$, this can be further simplified into
\begin{equation}
\label{equ:simple}
 Q = \frac{1}{var(\theta_1)} = H_{11} =   \sum_{l=1}^{N_s } 
f_l \cdot  \left(\frac{\partial \ln P_s (X_l;\Theta)} {\partial \theta_1}\right)^2.
\end{equation}

Only under the condition that all signal events have equal contribution to the measurement of $\theta_1$, which
means both local purity and $\partial \ln P_s (X_l;\Theta) / \partial \theta_1$ are constant,
the objective function can be reduced to the often misused form
\begin{equation}
\label{equ:significance}
 Q \propto F \cdot N_s = \frac {N_s^2}{N_s+N_b}.
\end{equation}
In general cases, maximizing $ N_s^2/(N_s+N_b) $  will lead to suboptimal solution
except for single-bin counting analysis.

\subsection{Case study: CP violation measurement in B decays}

We now apply the derived formalism to optimization problems in
CP violation study of neutral B decays, which  usually requires measuring the 
mixing induced CP asymmetry $S$ in time-dependent analysis.
 The  signal  p.d.f. is  
\begin{equation}
\label{equ:t_CP}
P_s(t,q;S) = \epsilon(t) \cdot e^{-\Gamma\cdot t} \cdot \left( 1+q\cdot (1-2\cdot\omega)\cdot S \cdot \sin(\Delta m \cdot t)\right )/I
\end{equation}
where the observables $(t,q)$ denote proper time and  B flavour tag, $I$ is a normalization factor independent of the 
 parameter $S$, $\epsilon(t)$ is an acceptance function describing the reconstruction efficiency as a function 
of proper time, $\Gamma$ and $\Delta m$ are parameters whose values are already measured in experiments, 
$\omega$ denotes the 
wrong tagging probability with a value in the range [0, 0.5].

Since $S$ is the only signal parameters in this case and the signal p.d.f. is a linear function of $S$, 
we can define the objective function following 
Eq.~(\ref{equ:simple}):
\begin{equation}
 Q = \frac{1}{var(S)} = H_{SS} 
   =  \sum_{l=1}^{N_s } f_l\cdot (1-2\cdot\omega_l)^2   \cdot 
      \left[ \frac{\sin(\Delta m\cdot t_l)}{ 1+q_l\cdot (1-2\cdot\omega_l)\cdot S \cdot \sin(\Delta m \cdot t_l) }  \right]^2 .
\end{equation}
It is worth noting that the shape of the acceptance function is irrelevant to evaluation of $Q$ and thus 
we can ignore it in the signal p.d.f. during the optimization process. 

So far we have not taken into account the effect of proper time resolution. Assuming a single Gaussian resolution model, the
signal p.d.f. becomes $P_s (t^{\prime},q;S) \otimes G(t-t^{\prime};\sigma_t,0) $, where $G$ is a 
Gaussian function 
with mean 0  and standard deviation $\sigma_t$. This effectively introduces an attenuation
factor into the sine term in Eq.~(\ref{equ:t_CP})  and changes  $Q$ into:
\begin{equation}
\label{equ:Q_CP}
 Q  =  \sum_{l=1}^{N_s } f_l\cdot (1-2\cdot\omega_l)^2   \cdot e^{-\left( \Delta m \cdot \sigma_{t,l}\right)^2 }  \cdot
      \left[ \frac{\sin(\Delta m\cdot t_l)}{ 1+q_l\cdot 
       (1-2\cdot\omega_l)\cdot S \cdot  e^{-\left( \Delta m \cdot \sigma_{t,l}\right)^2/2 } \cdot \sin(\Delta m \cdot t_l) }  
      \right]^2 .
\end{equation}
Qualitatively speaking, events with larger local purity, smaller wrong tagging probability and better proper time resolution
have bigger contribution to the sensitivity on $S$. This is consistent with intuitive expectation.

We stress again that only if all factors in the per event contribution to $Q$ are constant among  signal events,
the objective function $Q$ can be reduced to $ N_s^2/(N_s+N_b) $. Sometimes people try to maximize 
\begin{equation}
\label{equ:Q_close}
 Q = \frac{N_s^2}{N_s+N_b} \cdot (1-2\cdot\omega)^2   \cdot e^{-\left( \Delta m \cdot \sigma_t\right)^2 }  
\end{equation}
where $\omega$ and $\sigma_t$ are averaged over all accepted signal  events. This is closer to the objective
function we propose, but can still be misleading if any of the factors in the per event contribution
to $Q$ has big variation among signal events.
At hadron collider experiments, the proper time distributions of signal B decays and background B decays
are very different, with most background concentrating on $t\sim 0$ and signals having much longer lifetime.
The local purity is close to zero  at $t \sim 0$ and increases quickly with proper time. 
If one uses certain cuts to suppress  $t\sim 0$ background with the price of losing some signal events
with large $t$, this may help maximize the oversimplified
objective function in Eq.~(\ref{equ:Q_close}) but decreases the right one in  
Eq.~(\ref{equ:Q_CP}),
thus  leading to loss of sensitivity. 
This example demonstrates that a correct form of objective function should always be used in event selection optimization.

\section{Implication on choosing event variables}

Eq.~(\ref{equ:master}) and Eq.~(\ref{equ:simple}) clearly tell us that every signal event 
contributes to measurement sensitivity. Then a natural question to ask is: why do we 
perform selection at all? 

Indeed the best statistical precision is achieved when all possible information, including every
signal event and and every measured variable of each event,  is used in data fitting. 
However, it is impossible to do this in practice since the required signal and background model
will be too complicated to manage and the resulting losses in systematic accuracy will far exceed
the gains in statistical precision, even if we do not consider the requirement to reduce
event rates imposed by resource constraints. Therefore it is necessary to make some compromises
between systematic accuracy and statistical precision. 

According to how they are used in data analysis, event variables 
can be classified into three exclusive categories: 
fitting variables, binning variables and optimizeable variables.

Fitting variables are variables that are described in the probability density function and used in data fitting.
All variables needed for extraction of the physical parameters should be used as fitting variables. 
Variables that have power to separate signal from background and can be precisely modelled
should also be used as fitting variables. 
The requirement for fitting variables to be modellable guarantees that no 
big biases on physical measurements  will be induced due to inconsistency between 
fit model and data.

Binning variables are variables that are used to divide the data samples.
Variables that have power to separate signal from background but for which an accurate model
is difficult to find can   be used as binning variables. By dividing the data sample
into bins in a multi-dimensional space and treating each bin as an independent subsample, 
no events are dropped but local purity of signal events can still be enhanced. 
In practice only a small number of binning variables are allowed in an analysis, otherwise the number of 
subsamples will be too big and the complexity of the analysis will be  unmanageable. For this reason only
variables that will suffer big variation of local purity if they are used  as optimizieable variables should be
used as binning variables.

Optimizeable variables are  variables that are neither  fitting variables
nor binning variables. They can be used for background rejection.
Cutting on an optimizeable variable increases  the power of some signal events 
with the price of losing some low power signal events. 
Here the power of a signal event refers to  its contribution to the Hessian matrix elements.
If the gain is  greater than the loss, the measurement precision is improved.
The goal of optimization is to find the selection criteria on these variables
that minimize the measurement errors. That is why these variables  are called 
optimizeable variables. Any optimizeable variable can in principle be used for event 
selection. In practice it is difficult and unnecessary  to use all of them.
A set of   optimizeable  variables with the highest signal/background separation power
should be carefully chosen, for which the selection criteria are subject to optimization.

A selection optimization problem is  well defined only if all the three types of variables are 
meaningfully identified.
Before proceeding to define and maximize/minimize an objective function, make sure that:
the maximum information is used by including necessary variables as fitting and binning variables;
and the most powerful optimizeable variables are used to separate signal from background.

\section{Conclusions}

This paper presents a new approach to optimizing event selection for 
high energy physics measurements. Rather than using  sophisticated techniques 
to  optimize a poorly motivated goal,  
this approach aims to directly minimize the statistical uncertainty in the
physical measurements. A general formalism for quick and reliable error  estimation
is derived, and illustrated with the example of event selection optimization 
for CP violation measurement in B decays. The formalism not only makes direct
error optimization possible, but also has immediate implication on choosing
appropriate event variables for background rejection, data binning and data fitting.
In conclusion, event selection optimization for high energy physics measurements
can be and should be aimed at minimizing the errors on the physical results.

\section*{Acknowledgments}
The author would like to thank Vincenzo Vagnoni for pointing out a mistake in an earlier 
version of the paper.



\begin{thebibliography}{99}

\bibitem{statpat}
 I.~Narsky, 
 {\it StatPatternRecognition: A C++ Package for Statistical Analysis of High Energy Physics Data,}
 arXiv:physics/0507143.

\bibitem{tmva}
 A.~Hocker {\it et al.},
{\it TMVA - Toolkit for Multivariate Data Analysis,}
arXiv:physics/0703039.

\bibitem{open}
L.~Lyons,
{\it Open Statistical Issues in Particle Physics,}
arXiv:0811.1663.

\bibitem{Lyons}
L.~Lyons, {\it Statistics for Nuclear and Particle Physicists}, (Cambridge University Press, New York, 1986).

\bibitem{Cowan}
G.~Cowan, {\it Statistical Data Analysis}, (Oxford University Press, Oxford, 1998).


 \end{thebibliography}
\end{document}